# Hybrid quantum-classical convolutional neural network for phytoplankton classification

**Shangshang Shi**[†], **Zhimin Wang**[†,*], **Ruimin Shang, Yanan Li, Jiaxin Li, Guoqiang Zhong, and Yongjian Gu**[*]

Faculty of Information Science and Engineering, Ocean University of China, Qingdao, China

**\*Correspondence:** Zhimin Wang, wangzhimin@ouc.edu.cn; Yongjian Gu, yjgu@ouc.edu.cn

[†]These authors contribute equally to this work and share first authorship

**Abstract**

The taxonomic composition and abundance of phytoplankton, having direct impact on marine ecosystem dynamic and global environment change, are listed as essential ocean variables. Phytoplankton classification is very crucial for Phytoplankton analysis, but it is very difficult because of the huge amount and tiny volume of Phytoplankton. Machine learning is the principle way of performing phytoplankton image classification automatically. When carrying out large-scale research on the marine phytoplankton, the volume of data increases overwhelmingly and more powerful computational resources are required for the success of machine learning algorithms. Recently, quantum machine learning has emerged as the potential solution for large-scale data processing by harnessing the exponentially computational power of quantum computer. Here, for the first time, we demonstrate the feasibility of quantum deep neural networks for phytoplankton classification. Hybrid quantum-classical convolutional and residual neural networks are developed based on the classical architectures. These models make a proper balance between the limited function of the current quantum devices and the large size of phytoplankton images, which make it possible to perform phytoplankton classification on the near-term quantum computers. Better performance is obtained by the quantum-enhanced models against the classical counterparts. In particular, quantum models converge much faster than classical ones. The present quantum models are versatile, and can be applied for various tasks of image classification in the field of marine science.

## 1    Introduction

Phytoplankton is the most important primary producer in the aquatic ecosystem. Being the main supplier of dissolved oxygen in the ocean, phytoplankton plays a vital role in the energy flow, material circulation and information transmission in marine ecosystem (Barton et al., 2010; Gittings et al., 2018). Species composition and abundance of phytoplankton is one of the key factors of marine ecosystem dynamics, exerting a direct influence upon the global environment change. Therefore, much attention has been paid to the identification and classification of phytoplankton (Zheng et al., 2017; Pastore et al., 2020; Fuchs et al., 2022).

Nowadays, with the rapid development of imaging devices for phytoplankton (Owen et al., 2022), a huge number of phytoplankton images can be collected in a short time. Then it becomes impossible to classify and count these images using the traditional manual methods, i.e. expert-based methods.

In order to increase the efficiency of processing these images, machine learning methods has been introduced in, including the support vector machine (Hu et al., 2005; Sosik et al., 2007), random forest (Verikas et al., 2015; Faillettaz et al., 2016), k-nearest neighbor (Glüge et al., 2014), and artificial neural network (Mattei et al., 2018; Mattei et al., 2020). In particular, convolutional neural network (CNN), achieving state-of-the-art performance on image classification, becomes widely used in this field in recent years. A variety of CNN-based architectures were proposed to identify and classify phytoplankton with high efficiency and precision (Dai et al., 2017; Wang et al., 2018; Cui et al., 2018; Fuchs et al., 2022).

In order to perform large-scale research on the marine phytoplankton, more powerful computational resources are desired to guarantee the success of machine learning algorithms for processing the overwhelmingly increasing volume of data. On the other hand, there has been remarkable progress in the domain of quantum computing in recent years (Arute et al., 2019; Zhong et al., 2020; Bharti et al., 2022; Madsen et al., 2022). Quantum computing holds the promise of solving certain classically intractable problems (Preskill, 2018); quantum machine learning (QML) has emerged as the potential solution for large-scale data processing (Biamonte et al., 2017). In particular, there is a growing consensus that NISQ (noisy intermediate-scale quantum) devices may find advantageous applications in the near term (Callison et al., 2022), one of which is the quantum neural network (QNN) (Jeswal et al., 2019; Kwak et al., 2021). As a quantum analogue of classical neural network, QNN takes the parameterized quantum circuit (PQC) as a learning model (Benedetti et al., 2019), and can be extended naturally to quantum deep neural networks with the flexible multilayer architecture. Particularly, quantum convolutional neural network (QCNN) has got a lot of attention, and has demonstrated its success for processing both quantum data and classical data, including quantum many-body problems (Cong et al., 2019), identification of high-energy physics events (Chen et al., 2022), COVID-19 prediction (Houssein et al., 2022) and MNIST dataset classification (Oh et al., 2020). In general, there exits two architectures of QCNN, i.e. the fully quantum parameterized QCNN (Cong et al., 2019) and the hybrid quantum-classical CNN (Liu et al., 2021).

In this work, we present the potential of QCNN to perform phytoplankton classification. Considering the large size of phytoplankton images but the limited number of qubits and quantum operations available on current quantum devices, it is still unpractical to learn the images using fully quantum parameterized QCNN. Hence, we adopt the hybrid quantum-classical convolutional neural network (QCCNN) architecture to realize good multi-classification of phytoplankton dataset. QCCNN integrates the PQC into the classical CNN architecture by replacing the classical feature map with the quantum feature map. Therefore, QCCNN is friendly to current NISQ devices, in terms of both number of qubits and circuit depths, while retains important features of classical CNN, such as nonlinearity and scalability (Liu et al., 2021).

Furthermore, quantum neural networks would also suffer from the barren plateau problem (i.e. vanishing gradient) and the degradation problem (i.e. saturated accuracy with increasing depth) (Cong et al., 2019; Oh et al., 2020; Chen et al., 2022; Houssein et al., 2022). To address this issue, we further develop a QCCNN with residual structure, that is, a hybrid quantum-classical residual network (QCResNet), leveraging the residual architecture to enhance the performance of QCCNN.

The main contribution of the present work is as follows.

(1)   The feasibility of quantum neural networks to perform phytoplankton classification is demonstrated concretely for the first time. This work will definitely inspire more study of applying QML algorithms in marine science.

(2) Several specific architectures of QCCNN and QCResNet are developed, which achieve excellent performance on phytoplankton classification. Particularly, QCResNet architecture is proposed to increase the performance of QCCNN. These models are versatile, and can be applied directly for other tasks of image classification.

(3) Better performance on phytoplankton classification is obtained by QCCNN and QCResNet against the template CNN and ResNet models. The influence of the expressibility and entangling capability of PQCs on the performance of QCCNN is discussed.

The rest of the paper is organized as follows. Section 2 introduces preliminaries about the CNN, ResNet, and QNN. In section 3, architectures of QCCNN and QCResNet are discussed. Section 4 presents the experimental results about the performance of QCCNN and QCResNet as well as the dependence of QCCNN's performance on the feature of ansatz circuit. Conclusions are given in section 5.

## 2    Preliminaries

In the preliminaries, we present the basic architectures of CNN, ResNet and QNN with a certain degree of detail. These basic blocks will be used as framework to build the QCCNN and QCResNet models in the Methods section.

### 2.1    Convolutional neural network

CNN is one of the most successful tools in the area of computer vision (Gu et al., 2018). Since the first CNN known as Lenet was proposed (LeCun et al., 1998), there has been a series of architectures to improve the performance, such as Alexnet (Russakovsky et al., 2015), ZFNet (Zeiler et al., 2014), VGGNet (Simonyan et al., 2015), GoogleNet (Szegedy et al., 2015) and ResNet (He et al., 2016). These models are widely used in the field of image classification (Zuo et al., 2015; Lopes et al., 2017), object tracking (Nguyen et al., 2016; Li et al., 2017) and natural language processing (Kim et al., 2016).

The architecture of CNN is inspired by the visual perception mechanism of living creatures (Gu et al., 2018). CNN is comprised of three types of layers, which are the convolutional layers, pooling layers and fully-connected layers. The convolutional layers are applied in an attempt to extract useful features in the data which can be leveraged for classification purposes. Specifically, in a convolutional layer, a filter iteratively convolves local regions of the full input to produce feature maps, and each output feature contains information about different spatially-local patterns in the data (Henderson et al., 2020). When the convolutional layers are applied repeatedly, the increasingly abstract features are obtained, which capture the long-range dependencies within the data and are beneficial for the following classification or regression. In addition, the number of parameters in CNN is reduced substantially by the weight sharing between filters, increasing the computational efficiency of the model.

An example architecture of CNN is shown in Figure 1. This CNN consists of two convolutional layers, a max-pooling layer, and two fully connected layers. In this paper, this CNN is taken as the template framework to construct the QCCNN for phytoplankton classification.

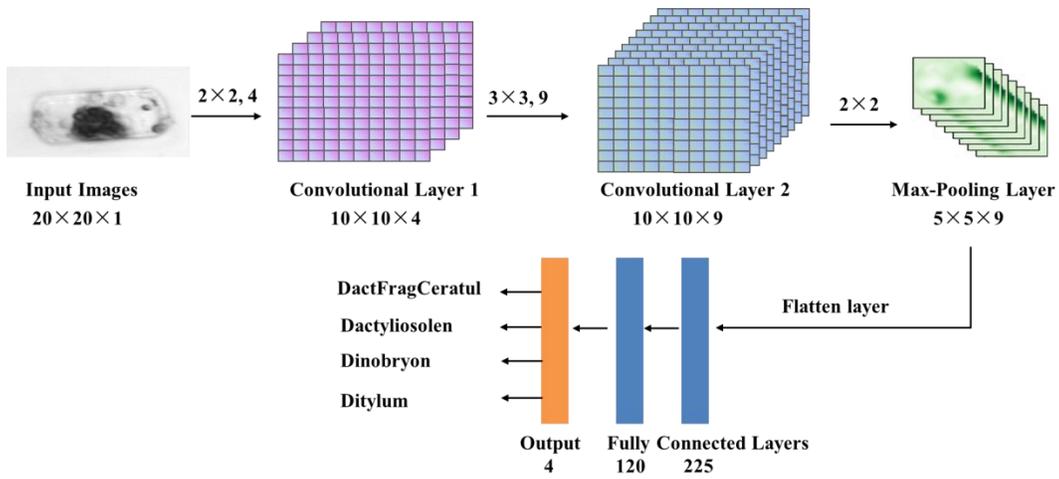

FIGURE 1 An example architecture of CNN that is used as the template to construct QCCNN.

## 2.2 Residual Network

In deep neural networks, network depth is of crucial importance for the performance of the network. However, when the network has large number of layers, two annoying problems emerge, i.e. the vanishing/exploding gradients and the saturation of accuracy. The residual learning framework was proposed to address the problems, making networks capable of having extremely deep architectures (He et al., 2016). The residual networks (ResNet) has been widely used in a wide range of applications, such as image recognition (Zagoruyko et al., 2016) and natural language processing (Conneau et al., 2016), showing compelling accuracy and nice convergence behaviors.

ResNet has the modularized architectures that stack residual units with the same connecting shape (He et al., 2016). The residual unit can be expressed as

$$\begin{aligned} y_l &= h(x_l) + F(x_l, W_l) \\ x_{l+1} &= f(y_l) \end{aligned}, \quad (1)$$

where $x_l$ and $x_{l+1}$ is the input and output of the $l$-th unit, and $F$ is the residual function. The central idea of ResNet is to learn the additive residual function $F$ with respect to $h(x_l)$. $h(x_l)$ can be implemented by the identity mapping, i.e. $h(x_l) = x_l$, or projection shortcut, i.e. the 1×1 convolution operation. Simply put, a CNN with residual architecture can be constructed by adding shortcut connections in the feedforward neural network.

Figure 2 shows an example of ResNet. It consists of two residual units, an adaptive average pooling layer and a fully connected layer. The 1×1 convolution operation is used in the shortcut connections. In this paper, this network is taken as the template framework to construct the QCResNet for phytoplankton classification.

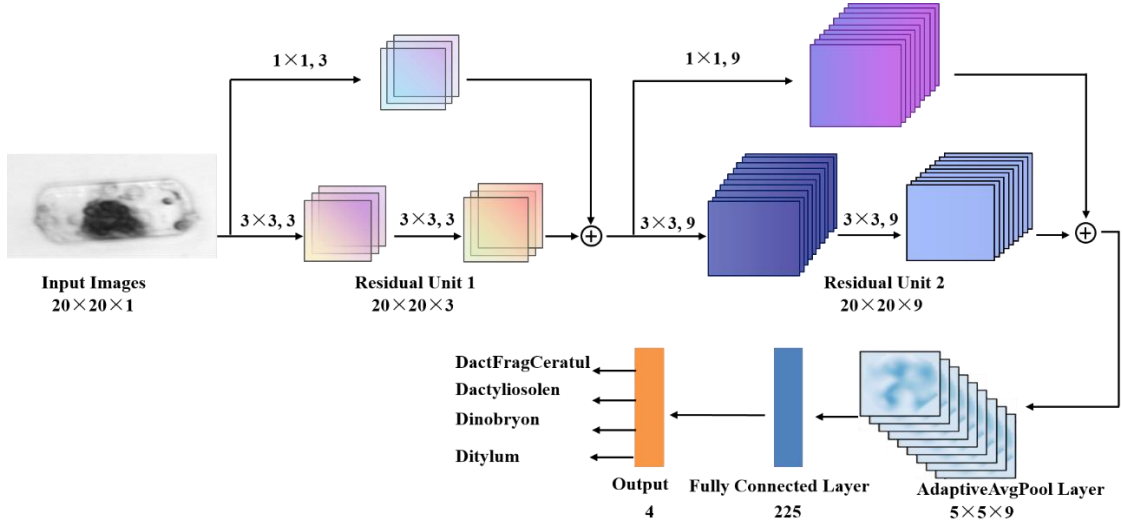

FIGURE 2 An example architecture of ResNet that is used as the template to construct QCResNet.

## 2.3 Quantum Neural Network

QNN belongs to the kind of variational quantum algorithms, which are the hybrid quantum-classical algorithms. In general, QNN is comprised of four parts, i.e. data encoding, ansatz performing forward transformation, quantum measurement and parameters optimization routine, as schematically shown in Figure 3. Note that the first three parts are implemented on the quantum device, while the optimization routine is carried out on the classical computer and feed the updated parameters back into quantum device.

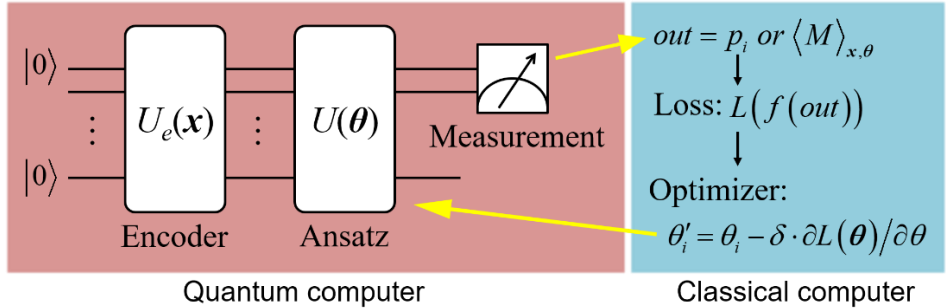

FIGURE 3 Architecture of QNN model. QNN is a quantum-classical hybrid algorithm. The forward transformation is implemented by the quantum computer and the parameters optimization is done by the classical computer.

Data encoding is to embed classical data into quantum states by applying a unitary transformation, i.e. $|x\rangle = U_e |0\rangle^{\otimes n}$ where $|x\rangle$ is proportional to the data vector $x$. Data encoding can be regarded as a quantum feature map, which is to map the data space to the quantum Hilbert space (Schuld et al., 2019). One of the prominent properties of QNN is introducing the quantum feature map into neural network that is extremely hard to simulate by classical resources (Havlíček et al., 2019). One of the most commonly used encoding method in QNN is the angle encoding. It embeds classical data into the rotation angles of the quantum rotation gates. For example, given a normalized data vector $x = (x_1, \ldots x_N)^T$ with $x_i \in [0,1]$, angle encoding can embed it as

$$R_y^{\otimes N}(x)|0\rangle^{\otimes N} = \bigotimes_{i=1}^{N}\left(\cos\frac{x_i}{2}|0\rangle + \sin\frac{x_i}{2}|1\rangle\right), \qquad (2)$$

where $R_y$ is the rotation gates about the $\hat{y}$ axes, i.e. $R_y(x_i) = \left[\cos\frac{x_i}{2}, -\sin\frac{x_i}{2}; \sin\frac{x_i}{2}, \cos\frac{x_i}{2}\right]$. More details about data encoding strategies can be found in (Hur et al., 2022).

Ansatz can be interpreted as a quantum analogue of feedforward neural network, using the quantum unitary transformation to implement the feature map of data. Ansatz is in fact a PQC with adjustable quantum gates. The adjustable parameters are optimized to approximate the target function that map features into different value domains representing different classes. So a proper structure of ansatz circuit plays a key role in specific learning tasks. Typically, QNN adopts the hardware-efficient ansatz, which use a limited set of quantum gates and particular qubit connection topology relating to the specific quantum devices on hand. The gate set usually contains three single-qubit gates and one two-qubit gates. An arbitrary single-qubit gate can be expressed as a combination of rotation gates about the $\hat{x}$, $\hat{y}$ and $\hat{z}$ axes. For example, using the X-Z decomposition, a single-qubit gate can be represented as

$$U_{1q}(\alpha, \beta, \gamma) = R_x(\alpha) R_z(\beta) R_x(\gamma), \qquad (3)$$

where $\alpha$, $\beta$, and $\gamma$ are the rotation angles. The two-qubit gates are used to produce entanglement between qubits. There are fixed two-qubit gates without adjustable parameters, such as the CNOT gate, and the one with adjustable parameters, such as the controlled $R_x(\theta)$ and $R_z(\theta)$ gates. A comprehensive discussion about the property of different ansatz circuit can be found in (Sim et al., 2019)

Quantum measurement is to output a value used as a prediction for the data. Measurement operation always corresponds to a Hermitian operator $M$ that can be decomposed as $M = \sum_i \lambda_i |i\rangle\langle i|$, where $\lambda_i$ is the $i$th eigenvalue and $|i\rangle$ is the corresponding eigenvector. After measurement, a quantum state $|\psi\rangle$ will collapse to one of the eigenstates $|i\rangle$ with a probability $p_i = |\langle i|\psi\rangle|^2$. Then, the expectation value of the measurement outcome is

$$\langle M \rangle = \sum_i \lambda_i \cdot p_i = \sum_i \lambda_i |\langle i|\psi\rangle|^2 . \qquad (4)$$

That is, the fundamental measurement outcome is the probabilities $\{p_i\}$ and the expectation $\langle M \rangle$. The most basic measurement in quantum computing is the computational basis measurement, also known as Pauli-$Z$ measurement, with the Hermitian operator

$$\sigma_z = (+1)|0\rangle\langle 0| + (-1)|1\rangle\langle 1|. \qquad (5)$$

When performing the $\sigma_z$ measurement, a qubit will collapse to the state $|0\rangle$ ($|1\rangle$) with the probability $p_0 = |\langle 0|\psi\rangle|^2$ ($p_1 = |\langle 1|\psi\rangle|^2$), from which we can read off the eigenvalues $+1$ or $-1$, respectively. The expectation value $\langle \sigma_z \rangle$ is a value in the range [-1, 1]. Because of the collapse principle of quantum measurement, in practice the probability $p_i$ and the expectation value are estimated by $s$ samples of measurement, where $s$ is known as the number of shots.

Optimization routine is used to update the parameters of ansatz circuit, i.e. the adjustable rotation angles of gates, based on the data. Optimizing the parameters is the process of minimizing the loss function $L(\theta)$ with respect to the parameter vector $\theta$. Almost the same loss functions as that used in the classical models, e.g. the mean squared error loss and the cross-entropy loss, can be applied in QNN. For instance, the multi-category cross-entropy loss can be expressed as

$$L(\theta) = -\frac{1}{N}\sum_{j=1}^{N}\sum_{c=1}^{C}\left[y_{jc} \cdot f\left(p_{i=c}\right)\right], \tag{6}$$

where $N$ is the batch size, $C$ is the number of categories, and $y_{jc} \in \{0,1\}$ is the class label. $p_{i=c}$ is the probability of measuring the eigenstates $|i\rangle$ corresponding to the category $c$. $f(\cdot)$ represents the post-processing of the measurement outcome, which is used to associate the outcome to the label $y_{jc}$.

Just like classical neural networks, the parameters can be updated based on the gradient of the loss function. Take the gradient descent method as example, the $i$th parameter $\theta_i$ is updated as

$$\theta_i' = \theta_i - \delta \cdot \frac{\partial L(\theta)}{\partial \theta_i}, \tag{7}$$

where $\delta$ is the learning rate. In quantum computing, there exists no backpropagation algorithm to directly calculate the gradient of loss. In practice, derivatives are evaluated using the difference method or the parameter shift rule (Wierichs et al., 2022) on the quantum devices. More details can be found in (Gyongyosi . et al., 2019).

## 3 Methods

### 3.1 Quantum-classical convolutional neural network

Using the CNN in Figure 1 as the template, the QCCNN can be obtained by replacing the convolutional layers with PQC. Since the template CNN has two convolutional layers, there are different replacement schemes. Figure 4 shows two possible QCCNN architectures. In Figure 4A, the QCCNN has a quantum convolutional layer and a classical convolutional layer, while Figure 4B has two quantum convolutional layers. Note that in order to take full advantage of quantum feature map to process the raw data, QCCNN always take the first convolutional layer as quantum one. Hereafter, the model of Figure 4A and Figure 4B is named as QCCNN-1 and QCCNN-2, respectively.

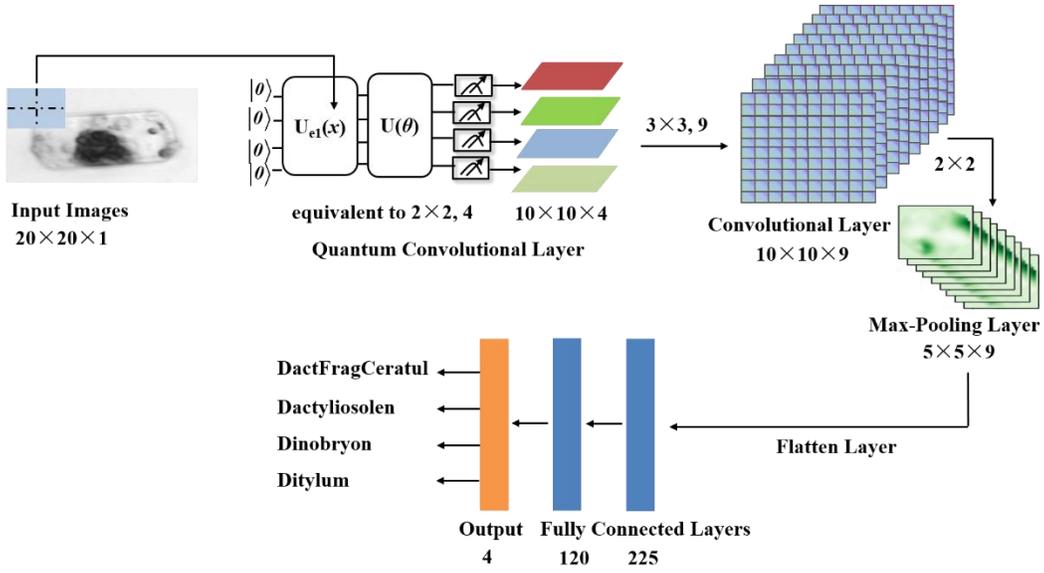

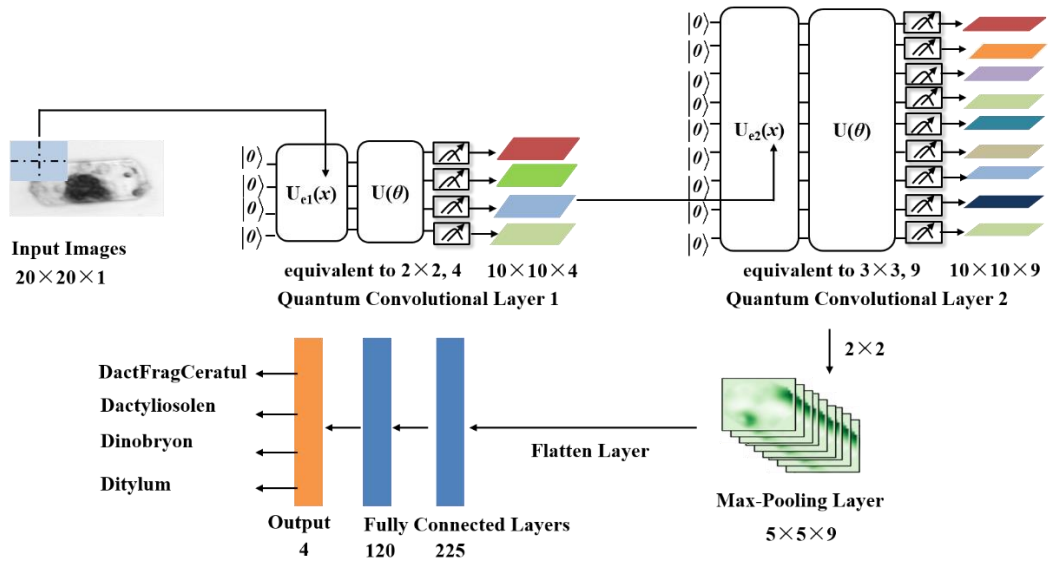

FIGURE 4 Architecture of the QCCNN with (A) one quantum and one classical convolutional layer (named as QCCNN-1) and (B) two quantum convolutional layers (named as QCCNN-2).

### 3.1.1 Quantum Convolutional Layer

The architecture of the first quantum convolution layer is shown in Figure 5. It is comprised of similar components as QNN, including the encoding circuit, ansatz circuit and quantum measurement.

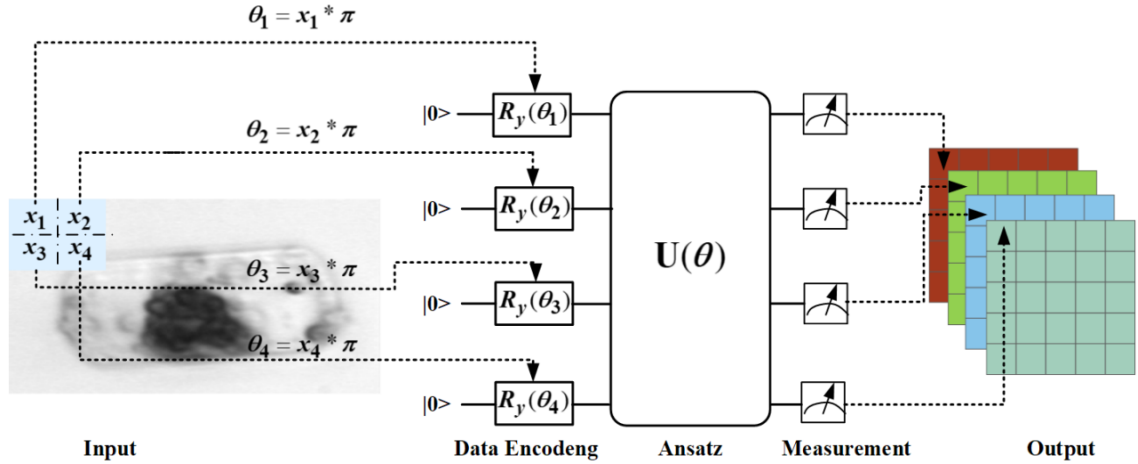

FIGURE 5 Architecture of the first quantum convolutional layer.

The window size of the filter is taken as 2×2, and the four elements are embedded using four qubits through four $R_y(\theta)$ gates. Two typical hardware-efficient ansatz are shown in Figure 6. They have different expressibility and entangling capability, which has large but still ambiguous impact on the performance of QNN. Figure 6A depicts the all-to-all configuration of two-qubit gates, which has the larger expressibility and entangling capability but the higher circuit complexity. By contrast, Figure 6B depicts the circuit-block configuration of two-qubit gates, which has the smaller expressibility and entangling capability but the lower circuit complexity (Sim et al., 2019). Of course, the expressibility and entangling capability of the ansatz can be increased by stacking the circuit as multi layers.

In quantum measurement, the four qubits are measured individually by the $\sigma_z$ operator. Then the four probabilities of each of the four qubits collapsing to $|0\rangle$ are taken as four feature channels for next layer as shown in Figure 5. Note that in the quantum convolutional layer there is not the activation function, and the nonlinearity comes from the process of data encoding and quantum measurement. This is the major different between QNN and CNN.

**A**

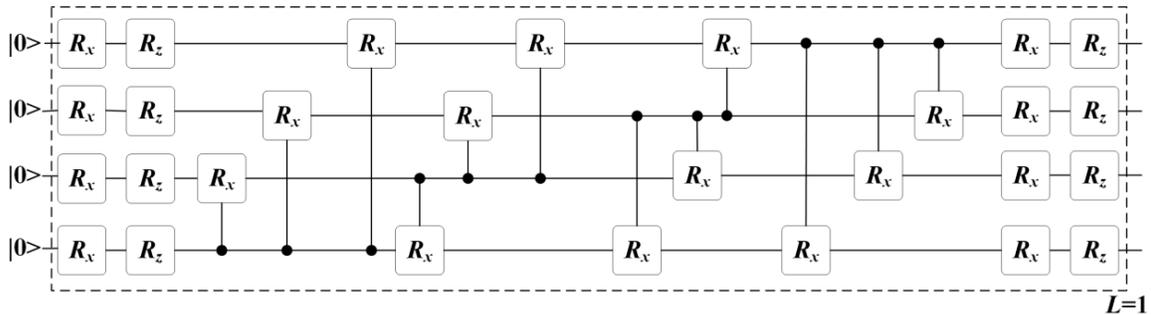

**B**

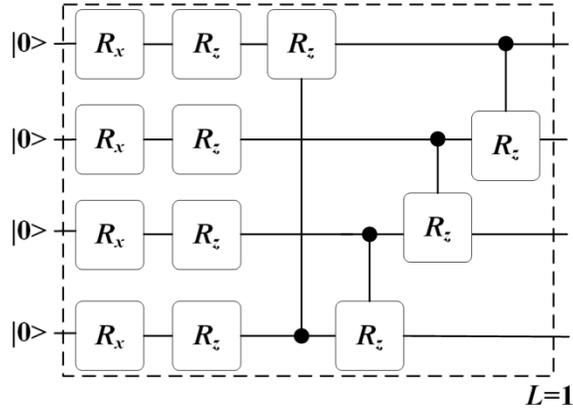

FIGURE 6 Two ansatz circuits with (A) all-to-all configuration and (B) circuit-block configuration of two-qubit gates. The circuit is used as one layer and several layers can be stacked to increase the expressibility and entangling capability of the circuit.

The architecture of the second quantum convolution layer is shown in Figure 7. In this case, the window size of the filter is taken as 3×3, and the nine elements are embedded using nine qubits through nine $R_y(\theta)$ rotation gates. The same ansatz circuits are used as shown in Figure 6. In quantum measurement, the nine qubits are measured individually by the $\sigma_z$ operator, and the nine probabilities of each of the nine qubits collapsing to $|0\rangle$ are taken as nine channels for next layer.

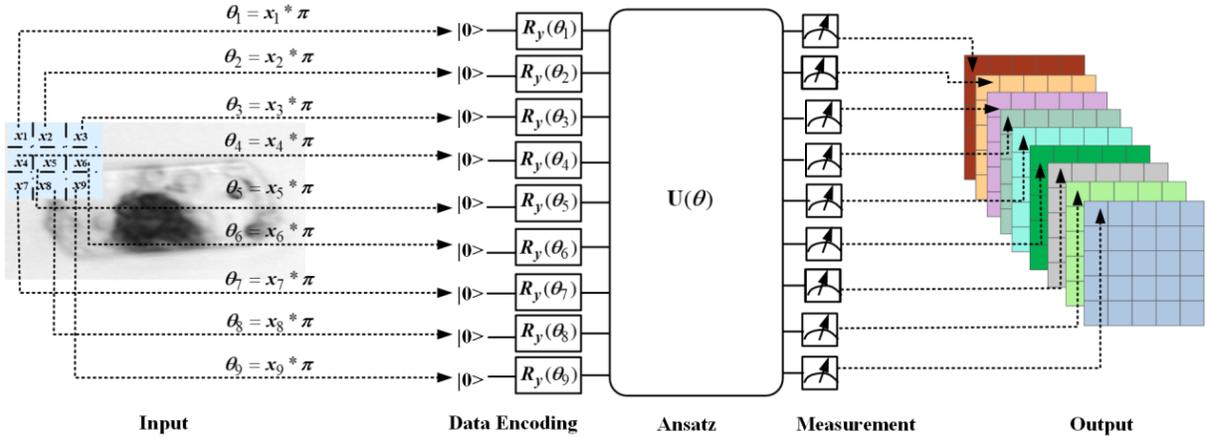

FIGURE 7 Architecture of the second quantum convolutional layer.

### 3.1.2 Classical Operations

The classical operations of QCCNN include the classical convolutional layers, the pooling layers, and the fully connected layers, which follows the typical operations of CNN. More specifically, in the convolutional layers, the window size is taken as 3×3. Activation function is the ReLu function. The Max Pooling layer is used to reduce the number of trainable parameters. At the end of QCCNN, two fully connected layers are always used to connect the convolutional and output layer.

### 3.2 Quantum Residual Network

Just as the way of constructing QCCNN, QCResNet can be obtained by replacing the convolutional layers with PQC in the template ResNet as shown in Figure 2. Since the template ResNet has two residual units, there are different replacement schemes. Figure 8 shows two possible QCResNet architectures. The first QCResNet as shown in Figure 8A has one quantum residual unit and the second one as shown in Figure 8B has two quantum residual unit. Hereafter, the model of Figure 8A and Figure 8B is named as QCResNet-1 and QCResNet-2, respectively.

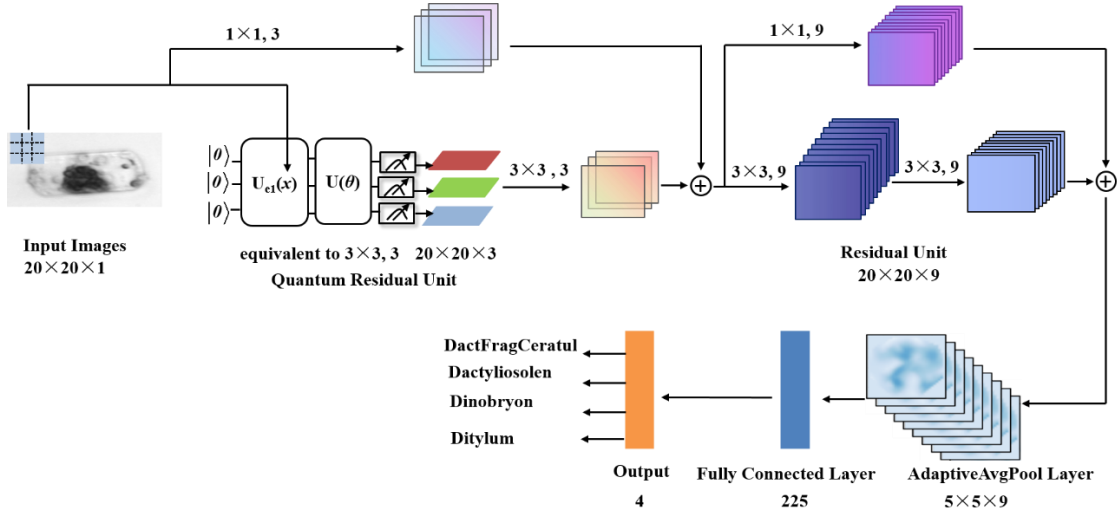

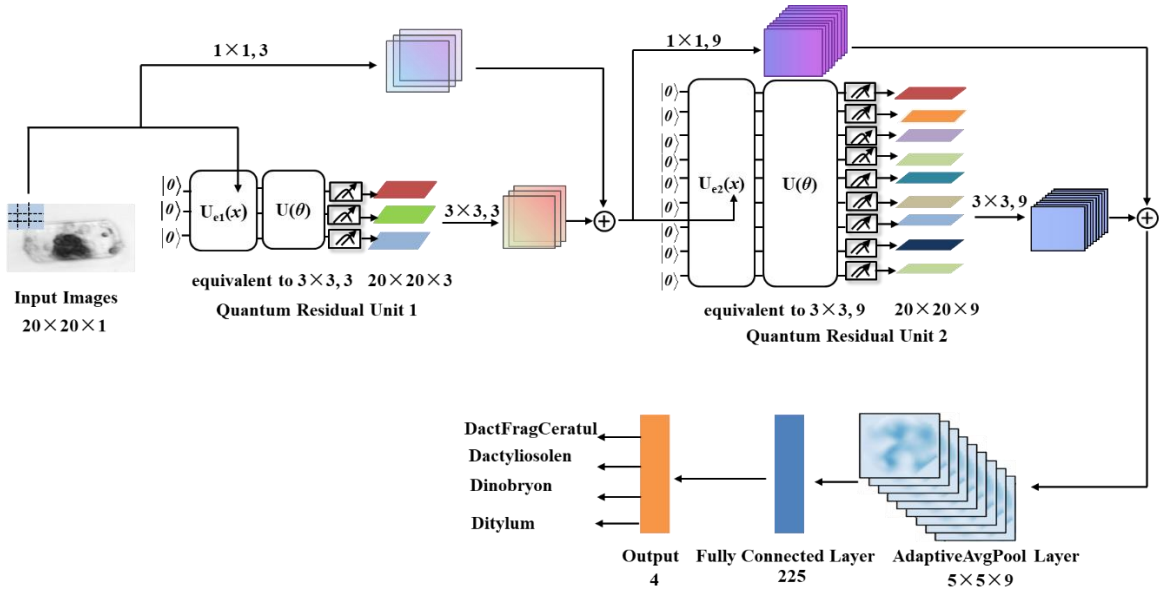

FIGURE 8 Architecture of the QCResNet with (A) one quantum residual unit (named as QCResNet-1) and (B) two quantum residual unit (named as QCResNet-2).

The operations of the first quantum residual unit are shown in Figure 9A, where the first convolutional layer is replaced with the quantum convolutional layer as shown in Figure 9B. The window size of the filter is taken as 3×3. In order to output three channels, the dense angle encoding method is applied, which use 3 qubits to embed 9 data elements. The ansatz circuit used in the quantum convolutional layer is the one as shown in Figure 6A. Finally, the three qubits are measured individually by the $\sigma_z$ operator, and the three probabilities of each of the three qubits collapsing to $|0\rangle$ are taken as three channels for next layer.

The architecture of the second quantum residual unit is almost the same as that of the first quantum residual unit except that the quantum convolutional layer used in the second quantum residual unit is the one in Figure 7.

**A**

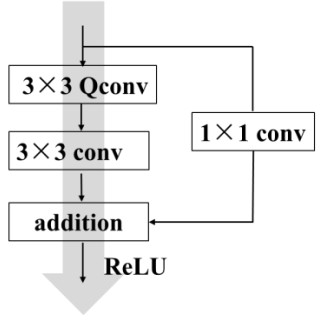

**B**

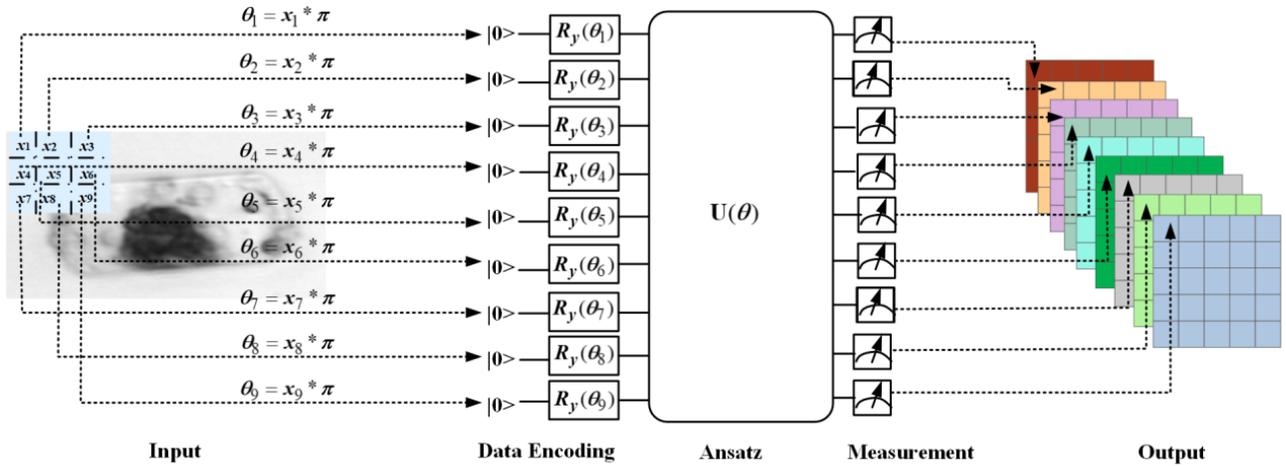

FIGURE 9 Operations of the first quantum residual unit (A), and the quantum convolutional layer used in the first quantum residual unit (B).

## 4   Results and discussion

### 4.1   Dataset and networks

The image dataset of phytoplankton used here were collected with a custom-built imaging-in-flow cytometer (Imaging FlowCytobot) by analyzing water from Woods Hole Harbor. All sampling was done between late fall and early spring in 2004 and 2005 (Sosik et al., 2007). The dataset comprises of 1200 images and contains 4 kinds of phytoplankton, i.e. *DactFragCeratul*, *Dactyliosolen*, *Dinobryon* and *Ditylum*. So each category contains 300 images. The 1200 images are divided unbiasedly into the training and testing subset with 600 images each. All images are normalized to 20×20 pixels.

Totally six neural networks are evaluated including the template CNN (see Figure 1), template ResNet (see Figure 2), QCCNN-1, QCCNN-2 (see Figure 4), QCResNet-1 and QCResNet-2 (see Figure 8). The ansatz used in the quantum convolutional layer of QCCNN and QCResNet is the circuit of Figure 6A. The cross-entropy function as shown in Eq. (6) is used as the loss function. The parameters in the quantum and classical layers are trained together, and they are updated based on the SGD method.

As discussed in the section of quantum measurement, the probability or expectation is estimated by repeating the measurement $s$ time, where $s$ is the number of shots. An appropriate number of shots need to be set a prior. To this end, a preliminary experiment is done using QCCNN-1. As shown in Figure 10, when $s = 1500$, QCCNN-1 achieves the best classification accuracy. So in the following experiments, 1500 shots are used in quantum measurement.

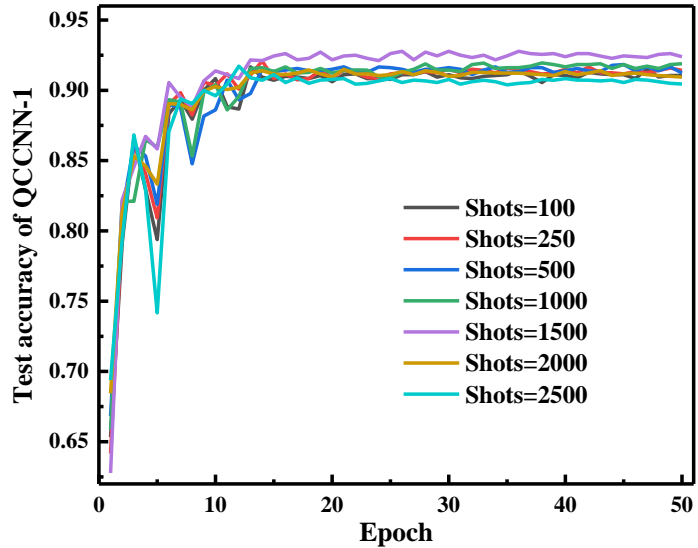

FIGURE 10 Test accuracy of QCCNN-1 on the phytoplankton classification using different number of shots in the process of quantum measurement.

## 4.2 Comparison of the performance

First, we compare the performance of the template CNN and QCCNN models. Figure 11 presents the variation of the test classification accuracy of the two models. As can be seen from the figure, the prominent feature of QCCNN is that it converges much faster than CNN. This should be due to the fact that QCCNN can represent much higher dimensional feature space and thus is capable of capturing more abstract information from data. Similar features have also been found in the quantum-inspired CNN (Shi et al., 2022).

The classification accuracy of QCCNN-1 is about 93.0%, which is close to that of CNN. But the accuracy fluctuation of QCCNN is much smaller, implying the good generalization of QCCNN. Furthermore, the accuracy of QCCNN-1 is higher than that of QCCNN-2, which implies that the replacement scheme of quantum convolutional layers has large impact on the performance of QCCNN. Specific replacement scheme should be trialed with regard to the learning task.

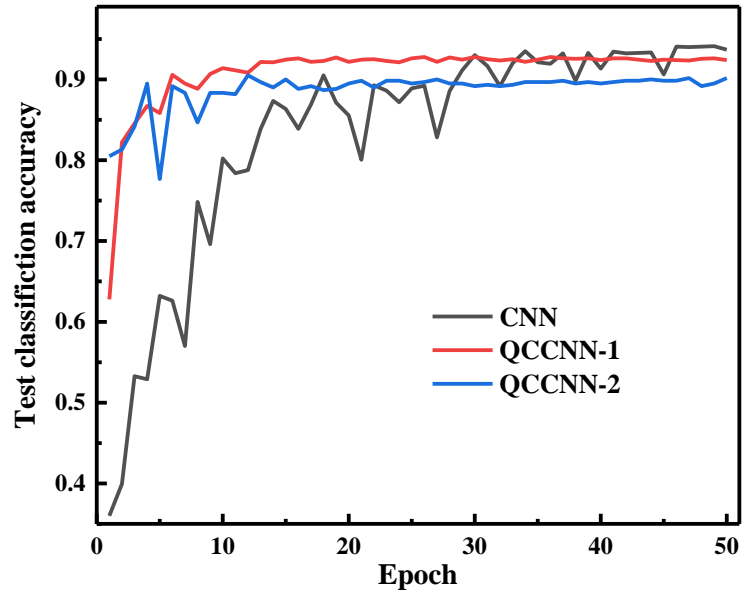

FIGURE 11 Curves of the test accuracy of QCCNNs and CNN for phytoplankton classification.

Next, we compare the performance of the template ResNet and QCResNet models. Figure 12 presents the variation of the test classification accuracy of the two models. Similar features can be found in QCResNet as that of QCCNN. As can be seen from the figure, QCResNet-1 converges faster than ResNet. The classification accuracy of QCResNet-1 is 93.2%, higher than the 91.9% of ResNet. However, QCResNet-2 has worse performance than ResNet, which again imply the large impact of the replacement scheme of quantum convolutional layers on the model performance. Although both QCResNet and ResNet present large fluctuation in the test accuracy curves, the fluctuation of QCResNet becomes smaller against ResNet after about 30 epoch. The large fluctuation should be due to the architecture of the template ResNet. We leave the optimization of ResNet for the future work.

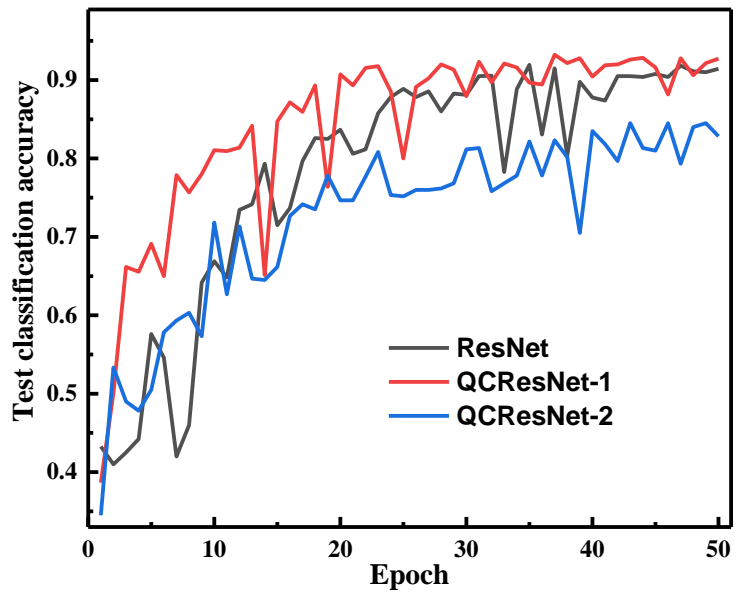

FIGURE 12 Curves of the test accuracy of QCResNet and ResNet for phytoplankton classification.

## 4.3 Influence of ansatz circuit for QCCNN

Quantum convolutional layer is the key component of both QCCNN and QResNet. Essentially, quantum convolutional layer takes the ansatz circuit, i.e. a PQC, as a filer to implement the forward transformation in CNN. Thus, features of the ansatz circuit should have direct influence on the performance of QCCNN and QResNet. Analyzing this relationship is beneficial to increase the performance of QNN models.

Ansatz circuit can be characterized quantificationally from the perspective of expressibility and entangling capability of PQC (Sim et al., 2019). Hence, there should be relations between the measures of PQC and the performance of the corresponding QCCNN. Below we use the QCCNN-1 as the basic model to exploit the dependence. Note that up to now QCCNN-1 use the circuit of Figure 6A as the ansatz. As discussed above, the circuit of Figure 6A has the larger expressibility and entangling capability, while the circuit of Figure 6B is lower but can be used in stack to increase its expressibility and entangling capability. Replace the ansatz of QCCNN-1 with multi-layers of the circuit of Figure 6B, then several versions of QCCNN-1 are obtained.

The classification accuracy of the five versions of QCCNN-1 is shown in Figure 13. As can be seen from the figure, the accuracy of QCCNN-1 using Figure 6A as the ansatz is higher than that using Figure 6B, which implies that higher expressibility and entangling capability of the ansatz circuit can indeed give rise to better performance of the QCCNN model.

However, for QCCNN-1 using multi-layers of Figure 6B as the ansatz, the accuracy does not always increase with the number of layer. The accuracy of QCCNN-1 with 2 layers is the highest, while that with 1, 3 and 4 layers is close. Note that the circuit using 4 layers of Figure 6B achieves the similar expressibility and entangling capability as that of Figure 6A. So this result implies that in addition to the property of expressibility and entangling capability, there are other influence factors. One is the number of trainable parameters. When increasing the number of layers, the expressibility and entangling capability increase, but the number of trainable parameters also increase. More parameters make the model harder to train and thus decrease its generalization, offsetting the positive effect of increasing the expressibility and entangling capability. The second factor would be the topological structure of the ansatz circuit. A quantificational way of characterizing the architecture of PQC and its correlation to the performance of the corresponding QCCNN need to be exploited in detail. We leave this for the future work.

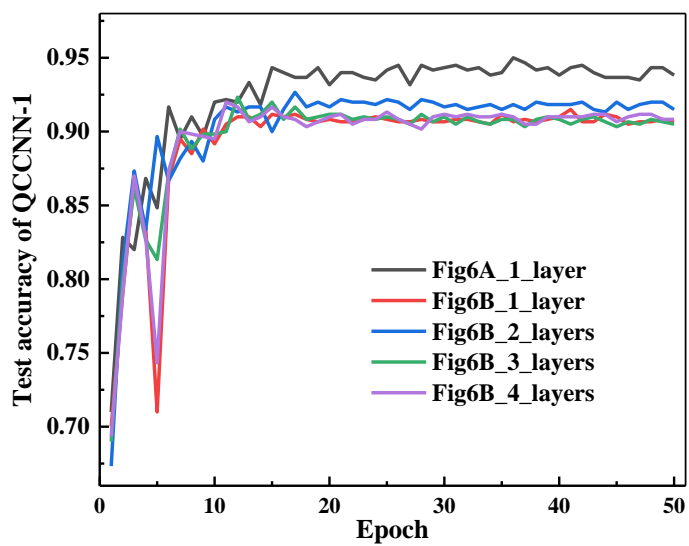

FIGURE 13 Classification accuracy of the five versions of QCCNN-1 using the circuit of Figure 6A and 1, 2, 3, 4 layers of Figure 6B as the ansatz.

## 5   Conclusion

In this work, we develop several hybrid quantum-classical convolutional and residual neural networks, and demonstrate their efficiency for phytoplankton classification. The QCCNN and QCResNet models are constructed by adding the quantum-enhanced forward transformation into the classical CNN and ResNet models. The hybrid architectures make a good balance between the limited function of the current NISQ devices and the large-size images of phytoplankton. Better performance of QCCNN and QCResNet is obtained against the classical counterpart. Especially, QCCNN and QCResNet can converge much faster than the classical models. Furthermore, we find that the performance of QCCNN and QCResNet depends on several factors, including the expressibility, entangling capability and topological structure of ansatz circuit and the number of training parameters. The model performance can be increased by taking into account all these factors. The present QCCNN and QCResNet models can be expanded easily for other tasks of image classification. In future, we will optimize the architecture of QCCNN and QCResNet from the aspect of both quantum and classical convolutional layers, and exploit the applications of the quantum-enhanced models in other tasks of marine science.

**Data and Code availability**

The data and codes are available upon request from the authors.

**Conflict of interest**

The authors declare that they have no known competing financial interests or personal relationships that could have appeared to influence the work reported in this paper.

**Author contributions**

SS and ZW developed the algorithms and wrote the first draft. SS, RS, YL and JL wrote the codes and carried out the numerical experiments. YG, GZ and ZW planned and designed the project. All authors discussed the results and reviewed the manuscript.

**Funding**

The present work is supported by the Natural Science Foundation of Shandong Province of China (ZR2021ZD19) and the National Natural Science Foundation of China (12005212).

**Acknowledgments**

We are grateful to the support of computational resources from the Marine Big Data Center of Institute for Advanced Ocean Study of Ocean University of China.**Reference**

Arute, F., Arya, K., Babbush, R., Bacon, D., Bardin, J. C., Barends, R., et al. (2019). Quantum supremacy using a programmable superconducting processor. *Nature*, 574(7779), 505-510. doi:10.1038/s41586-019-1666-5

Barton, A. D., Dutkiewicz, S., Flierl, G., Bragg, J., and Follows, M. J. (2010). Patterns of diversity in marine phytoplankton. *Science*, 327(5972), 1509-1511. doi: 10.1126/science.1184961

Benedetti, M., Lloyd, E., Sack, S., and Fiorentini, M. (2019). Parameterized quantum circuits as machine learning models. *Quantum Science and Technology*, 4(4), 043001. doi: 10.1088/2058-9565/ab4eb5

Bharti, K., Cervera-Lierta, A., Kyaw, T. H., Haug, T., Alperin-Lea, S., Anand, A., et al. (2022). Noisy intermediate-scale quantum algorithms. *Reviews of Modern Physics*, 94(1), 015004. doi:10.1103/RevModPhys.94.015004

Biamonte, J., Wittek, P., Pancotti, N., Rebentrost, P., Wiebe, N., and Lloyd, S. (2017). Quantum machine learning. *Nature*, 549, 195. doi:10.1038/nature23474

Callison, A., and Chancellor, N. (2022). Hybrid quantum-classical algorithms in the noisy intermediate-scale quantum era and beyond. *Physical Review A*, 106(1), 010101. doi: 10.1103/PhysRevA.106.010101

Chen, S. Y. C., Wei, T. C., Zhang, C., Yu, H., and Yoo, S. (2022). Quantum convolutional neural networks for high energy physics data analysis. *Physical Review Research*. 4(1), 013231. doi:10.1103/PhysRevResearch.4.013231

Cong, I., Choi, S., and Lukin, M. D. (2019). Quantum convolutional neural networks. *Nature Physics*. 15(12), 1273-1278. doi: 10.1038/s41567-019-0648-8

Conneau, A., Schwenk, H., Barrault, L., and Lecun, Y. (2016). Very deep convolutional networks for natural language processing. *arXiv preprint* arXiv:1606.01781, 2(1)

Cui, J., Wei, B., Wang, C., Yu, Z., Zheng, H., Zheng, B., and Yang, H. (2018). Texture and shape information fusion of convolutional neural network for plankton image classification. *2018 OCEANS - MTS/IEEE Kobe Techno-Oceans (OTO)*. Kobe, Japan, 2018, pp. 1-5, doi: 10.1109/OCEANSKOBE.2018.8559156

Dai, J., Yu, Z., Zheng, H., Zheng, B., and Wang, N. (2017). A hybrid convolutional neural network for plankton classification. In *Computer Vision–ACCV 2016 Workshops: ACCV 2016 International Workshops*, 102-114. doi : 10.1007/978-3-319-54526-4_8

Faillettaz, R., Picheral, M., Luo, J. Y., Guigand, C., Cowen, R. K., and Irisson, J. O. (2016). Imperfect automatic image classification successfully describes plankton distribution patterns. *Methods in Oceanography*, 15, 60-77. doi:10.1016/j.mio.2016.04.00317

Fuchs, R., Thyssen, M., Creach, V., Dugenne, M., Izard, L., Latimier, M., et al. (2022). Automatic recognition of flow cytometric phytoplankton functional groups using convolutional neural networks. *Limnology and Oceanography: Methods*. doi: 10.1002/lom3.10493

Gittings, J. A., Raitsos, D. E., Krokos, G., and Hoteit, I. (2018). Impacts of warming on phytoplankton abundance and phenology in a typical tropical marine ecosystem. *Scientific reports*. 8(1), 1-12. doi: 10.1038/s41598-018-20560-5

Glüge, S., Pomati, F., Albert, C., Kauf, P., and Ott, T. (2014). The Challange of Clustering Flow Cytometry Data from Phytoplankton in Lakes. In: *Mladenov, V.M., Ivanov, P.C. (eds) Nonlinear Dynamics of Electronic Systems. NDES 2014. Communications in Computer and Information Science, vol 438.* Springer, Cham. doi: 10.1007/978-3-319-08672-9_45

Gu, J., Wang, Z., Kuen, J., Ma, L., Shahroudy, A., Shuai, B., et al. (2018). Recent advances in convolutional neural networks. *Pattern recognition*, 77, 354-377. doi: 10.1016/j.patcog.2017.10.013

Gyongyosi, L., and Imre, S. (2019). Training optimization for gate-model quantum neural networks. *Scientific Reports*, 9(1), 1-19. doi: 10.1038/s41598-019-48892-w

He, K., Zhang, X., Ren, S., and Sun, J. (2016). Deep residual learning for image recognition. In *Proceedings of the IEEE conference on computer vision and pattern recognition*, 770-778. doi: 10.1109/CVPR.2016.9z

Henderson, M., Shakya, S., Pradhan, S., and Cook, T. (2020). Quanvolutional neural networks: powering image recognition with quantum circuits. *Quantum Machine Intelligence*, 2(1), 1-9. doi: /10.1007/s42484-020-00012-y

Havlíček, V., Córcoles, A. D., Temme, K., Harrow, A. W., Kandala, A., Chow, J. M., Gambetta, J. M. (2017). Supervised learning with quantum-enhanced feature spaces. *Nature*, 567, 209. https://doi.org/10.1038/s41586-019-0980-2

Houssein, E. H., Abohashima, Z., Elhoseny, M., and Mohamed, W. M. (2022). Hybrid quantum-classical conv*olutional* neural network model for COVID-19 prediction using chest X-ray images. *Journal of Computational Design and Engineering*, 9(2), 343-363. doi: 10.1093/jcde/qwac003

Hu, Q., and Davis, C. (2005). Automatic plankton image recognition with co-occurrence matrices and support vector machine. *Marine Ecology Progress Series*. 295, 21-31. doi:10.3354/meps295021

Hur, T., Kim, L., and Park, D. K. (2022). Quantum convolutional neural network for classical data classification. *Quantum Machine Intelligence*, 4(1), 1-18. doi: 10.1007/s42484-021-00061-x

Jeswal, S. K., and Chakraverty, S. (2019). Recent developments and applications in quantum neural network: a review. *Archives of Computational Methods in Engineering*. 26(4), 793-807. doi: 10.1007/s11831-018-9269-0

Kim, Y., Jernite, Y., Sontag, D., and Rush, A. M. (2016). Character-aware neural language models. In *Thirtieth AAAI conference on artificial intelligence*. doi: 10.1609/aaai.v30i1.10362

Kwak, Y., Yun, W. J., Jung, S., and Kim, J. (2021). Quantum neural networks: Concepts, applications, and challenges. In *2021 Twelfth International Conference on Ubiquitous and Future Networks (ICUFN)*, 413-416. doi: 10.1109/ICUFN49451.2021.9528698

LeCun, Y., Bottou, L., Bengio, Y., and Haffner, P. (1998). Gradient-based learning applied to document recognition. *Proceedings of the IEEE*, 86(11), 2278-2324. doi: 10.1109/5.726791
18